\documentclass[sigconf]{acmart}

\usepackage{booktabs} 
\usepackage{url}
\usepackage{color}
\usepackage{soul}
\usepackage{booktabs}
\usepackage{graphicx}
\usepackage{multirow}
\usepackage{mathrsfs}
\usepackage{amssymb}
\usepackage{amsbsy}
\usepackage{amsmath}
\usepackage{tcolorbox}
\usepackage{balance}
\usepackage{microtype}

\setcopyright{rightsretained}

\acmDOI{XXX}

\acmISBN{XXX}

\acmConference[ICSE'18 NIER]{ICSE}{May-June 2018}{Gothenburg, Sweden}
\acmYear{2018}
\copyrightyear{2018}

\acmPrice{15.00}

\begin{document}
\title{Explainable Software Analytics}

\author{Hoa Khanh Dam}
\affiliation{%
	\institution{University of Wollongong, Australia}
}
\email{hoa@uow.edu.au}

\author{Truyen Tran}
\affiliation{%
	\institution{Deakin University, Australia}
}
\email{truyen.tran@deakin.edu.au}

\author{Aditya Ghose}
\affiliation{%
	\institution{University of Wollongong, Australia}
}
\email{aditya@uow.edu.au}

%


\begin{abstract}

Software analytics has been the subject of considerable recent attention but is yet to receive significant industry traction. One of the key reasons is that software practitioners are reluctant to trust predictions produced by the analytics machinery without understanding the rationale for those predictions. While complex models such as deep learning and ensemble methods improve predictive performance, they have limited explainability. In this paper, we argue that making software analytics models explainable to software practitioners is as \emph{important} as achieving accurate predictions. Explainability should therefore be a key measure for evaluating software analytics models. We envision that explainability will be a key driver for developing software analytics models that are useful in practice. We outline a research roadmap for this space, building on social science, explainable artificial intelligence and software engineering.

\end{abstract}

%

%


\maketitle

\section{Introduction}

Software analytics has in recent times become the focus of considerable research attention.
The vast majority of software analytics methods leverage machine learning algorithms to build prediction models for various software engineering tasks such as defect prediction, effort estimation, API recommendation, and risk prediction. Prediction accuracy 
(assessed using measures such as precision, recall, F-measure, Mean Absolute Error and similar)
is currently the dominant criterion for evaluating predictive models in the software analytics literature. 
The pressure to maximize predictive accuracy has led to the use of powerful and complex models such as SVMs, ensemble methods and deep neural networks.
Those models are however considered as ``black box'' models, thus it is difficult for software practitioners to understand them and interpret their predictions.

The lack of explainability results in the lack of \emph{trust},
leading to industry reluctance to adopt or deploy software analytics.
If software practitioners do not understand a model's predictions, they would \emph{not} blindly trust those predictions,
nor commit project resources or time to act on those predictions.
High predictive performance during testing may establish some degree of trust in the model. However, those test results may not hold when the model is deployed and run ``in the wild''. After all, the test data is already known  to the model designer, while the ``future'' data is unknown and can be completely
different from the test data, potentially making the predictive machinery far less effective.
Trust is especially important when the model produces predictions that are different from the software practitioner's expectations
(e.g., generating alerts for parts of the code-base that developers expect to be ``clean'').
Explainability is not only a pre-requisite for practitioner trust, but also potentially a trigger for new insights in the minds of practitioners.
For example, a developer would want to understand why a defect prediction model suggests that a particular source file is defective so that they can fix the defect.
In other words, merely flagging a file as defective is often not good enough. Developers need a trace and a justification of such predictions in order to generate appropriate fixes.

We believe that making predictive models explainable to software practitioners is as \emph{important} as achieving accurate predictions. Explainability should be an important measure for evaluating software analytics models. We envision that explainability will be a key driver for developing
software analytics models that actually deliver value in practice.
In this paper, we raise a number of research questions that will be critical to the success of software analytics in practice:

\begin{enumerate}
  \item What forms a (good) explanation in software engineering tasks?
  \item How do we build explainable software analytics models and how do we generate explanations from them?
  \item How might we evaluate explainability of software analytics models?
\end{enumerate}

Decades of research in social sciences have established a strong understanding of how humans explain decisions and behaviour to each other \cite{Miller17a}. This body of knowledge will be useful for developing software analytics models that are truly capable of providing explanations to human engineers. In addition, \emph{explainable artificial intelligence} (XAI) has recently become an important field in AI. AI researchers have started looking into developing new machine learning systems which will be capable of explaining their decisions and describing their strengths and weaknesses
in a way that human users can understand (which in turn engenders {\em trust} that these systems will generate reliable results in future usage scenarios).
In order to address these research questions,
we believe that the software engineering community
needs to adopt an inter-disciplinary approach that brings together
existing work from social sciences, state-of-the-art work in AI, and the existing domain knowledge of software engineering.
In the remainder of this paper, we will discuss these research directions in greater detail.



\section{What is a good explanation in software engineering?}
Substantial research in philosophy, social science, psychology,
and cognitive science has sought to understand what constitutes explainability and how explanations might be presented in a form that would be easily understood (and thence, accepted) by humans.
There are various definitions of explainability, but for software analytics we propose to adopt a simple definition: \emph{explainability or interpretability of a model measures the degree to which a human observer can understand the reasons behind a decision (e.g. a prediction) made by the model}. There are two distinct ways of achieving explainability: (i) making the entire decision making process transparent and comprehensible; and (ii) explicitly providing explanation for each decision. The former is also known as global explainability, while the latter refers to local explainability (i.e. knowing the reasons for a specific prediction) or post-hoc interpretability \cite{Lipton16}.

Many philosophical theories of explanation (e.g. \cite{Salmon84}) consider explanations as presentation of causes, while other studies (e.g. \cite{Lipton1990}) view explanations as being contrastive. We adapted the model of explanation developed in \cite{VanBouwel2002}, which combines both of these views by considering explanations as answers to four types of why-questions:

\begin{itemize}
   \item (plain fact) \emph{``Why does object a have property P?''} \\
    Example: Why is file A defective? Why is this sequence of API usage recommended?

    \item (P-contrast) \emph{``Why does object a have property P, rather than property P'?''} \\
  Example: Why is file A defective rather than clean?

  \item (O-contrast) \emph{``Why does object a have property P, while object b has property P'?''} \\
  Example: Why is file A defective, while file B is clean?

  \item (T-contrast) \emph{``Why does object a have property P at time t, but property P' at
time t'?''}  \\
   Example:
	Why was the issue not classified as delayed at the beginning, but was subsequently classified as delayed three days later?
\end{itemize}

An explanation of a decision (which can involve the statement of a plain fact) can take the form of a causal chain of events that lead to the decision.
Identify the complete set of causes is however difficult, and most
machine learning algorithms offer correlations as opposed to categorical statements of causation.
The three contrastive questions constitute a contrastive explanation, which is often easier to produce than answering a plain fact question since contrastive explanations focus on only the differences between two cases. Since software engineering is task-oriented, explanations in software engineering tasks should be viewed from the pragmatic explanatory pluralism perspective \cite{DeWinter2010}. This means that there are multiple legitimate types of explanations in software engineering. Software engineers have
have different preferences for the types of explanations they are willing to accept,
depending on their interest, expertise and motivation.


%
%




\section{Explainable software analytics models}

One effective way of achieving explainability is making the model transparent to software practitioners. In this case, the model itself is an explanation of how a decision is made.
Transparency could be achieved at three levels:
the entire model, each individual component of the model (e.g. input features or parameters), and the learning algorithm \cite{Lipton16}.

\begin{figure}[h]
	\centering
	\includegraphics[width=\linewidth]{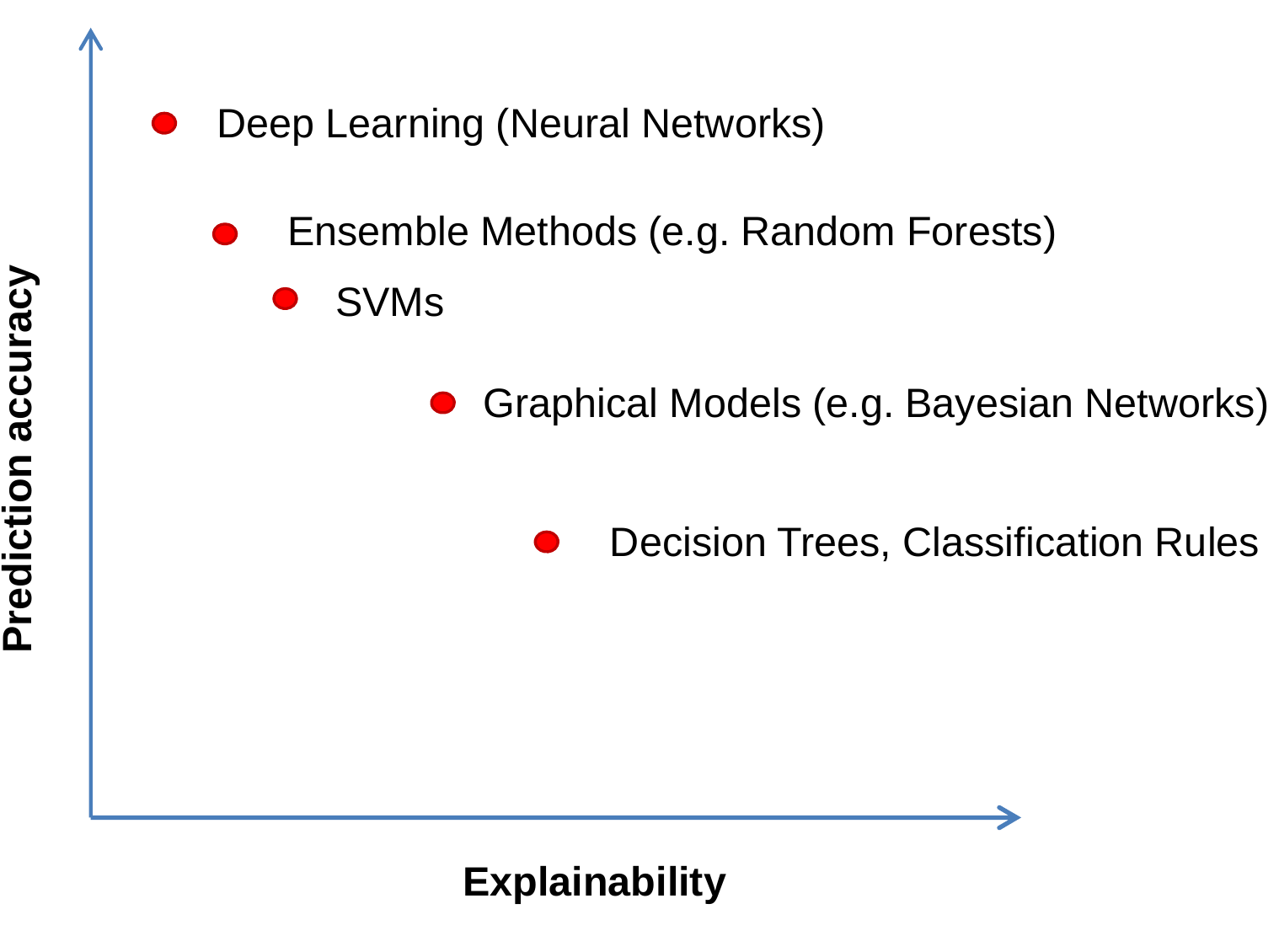}
	\caption{Prediction accuracy versus explainability}
	\label{fig:tradeoff}
\end{figure}

The Occam's Razor principle of parsimony suggests that an software analytics model needs to be expressed in a simple way that is  easy for software practitioners to interpret \cite{DBLP:books/Menzies14}. Simple models such as decision tree, classification rules, and linear regression tend to be more transparent than complex models such as deep neural networks, SVMs or ensemble methods (see Figure \ref{fig:tradeoff}). Decision trees have a graphical structure which facilitate the visualization of a decision making process. They contain only a subset of features and thus help the users focus on the most relevant features. The hierarchical structure also reflect the relative importance of different features: the lower the depth of a feature, the more relevant the feature is for classification. Classification rules, which can be derived from a decision tree, are considered to be one of the most interpretable classification models in the form of sparse decision lists. Those rules have the form of IF (conditions) THEN (class) statements, each of which specifies a subset of features and a corresponding predicted outcome of interest. Thus, those rules naturally explain the reasons for each prediction made by the model. The learning algorithm of those simple models are also easy to comprehend, while powerful methods such as deep learning algorithms are complex and lack of transparency.

Each individual component of an explainable model (e.g. parameters and features) also needs to provide an intuitive explanation. For example, each node in a decision tree may refer to an explanation, e.g. the normalized cyclomatic complexity of a source file is less than 0.05. This would exclude features that are anonymized (since they contain sensitive information such as the number of defects and the development time). Highly engineered features (e.g. TF-IDF that are commonly used) are also difficult to interpret. Parameters of a linear model could represent strengths of association between each feature and the outcome.

Simple linear models are however not always interpretable. Large decision trees with many splits are difficult to comprehend. Feature selection and pre-processing can have an impact. For example, associations between the normalized cyclomatic complexity and defectiveness might be positive or negative, depending on whether the feature set includes other features such as lines of code or the depth of inheritance trees. 

Future research in software analytics should explicitly address the \emph{trade-off between explainability and prediction accuracy}. It is very hard to define in advance which explainability is suitable to a software engineer. Hence, a multi-objective approach based on Pareto dominance would be more suitable to sufficiently address this trade-off. For example, recent work \cite{Wisam17} has employed an evolutionary search for a model to predict issue resolution time. The search is guided simultaneously by two contrasting objectives: maximizing the accuracy of the prediction model while minimizing its complexity. 

The use of simple models improves explainability but requires a sacrifice for accuracy. An important line of research is finding methods to tune those simple models in order to improve their predictive performance. For example, recent work \cite{Fu:2017:EOH} has demonstrated that evolutionary search techniques can also be used to fine tune SVM such that it achieved better predictive performance than a deep learning method in predicting links between questions in Stack Overflow. Another important direction of research is making black-box models more explainable, which enable human engineers to understand and  appropriately trust the decisions made by software analytics models. In the next section, we discuss a few research directions in making deep learning models more explainable.

%

%
%
%
%
%
%
%
%
%
%
%
%
%
%
%
%
%
\section{Explanations for deep models}

Deep learning, an AI methodology for training deep neural networks,
has found use in a number of recent developments in software analytics.
The power of deep learning mainly comes from architectural flexibility,
which enables design of neural nets to fit almost any complex
software structures (e.g., those found in `Big Code' \cite{raychev2015predicting}).
Recent examples include code/text sequences \cite{dam2016deepsoft},
AST \cite{li2017software}, API learning \cite{gu2016deep},
code generation \cite{rabinovich2017abstract} and project dependency networks \cite{pham2017column}.

This flexibility comes at the cost of explainability.
Deep learning methods were primarily developed for \emph{distant credit
assignments} through a long chain of nonlinear transformations.
Thus understanding model's decision becomes a great challenge. Even
when it is possible to understand, great care must be taken
to interpret the results because deep networks can fit noise and
find arbitrary hypotheses \cite{zhang2017understanding}.


To address this lack of explainability in deep learning, there has been
a growing effort to turn a black--box neural net into a \emph{clear--box}, which
we discuss below.

\subsection{Seeing through the black-box}
This strategy involves looking into the inner working of existing neural nets.
A deep neural net is constructed by stacking primitive components
known as neurons. A neuron is a simple non-linear function of several inputs.
The entire network of neuron forms a computational graph, which
allows informative signals from data to pass through and reach output nodes.
Likewise, training signals from labels back-propagate in the
reverse directions, assigning credits to each neuron and neuron pair
along the way.

A popular way to analyse a deep net
is through \emph{visualization} of its neuron activations and weights.
This often provides an intuitive explanation of
how the network works internally. Well-known examples include visualizing
token embedding in 2D \cite{choetkiertikul2016deep}, feature maps in CNN \cite{simonyan2013deep}
and recurrent activation patterns in RNN \cite{karpathy2015visualizing}.
These techniques can also  estimate \emph{feature importance} \cite{ibrahim2013comparison}
e.g., by differentiating the function $f(x,W)$ against feature $x_i$.

\subsection{Designing explainable architectures}

An effective strategy is to design new architectures that self-explain
decision making at each step.
One of the most useful mechanisms is through \emph{attention},
in which model components compete to contribute to an outcome (e.g., see \cite{bahdanau2015neural}).
Examples of components include code token in a sequence, a node
in an AST, or a function in a class.


Another method is \emph{coupling neural nets with natural language processing} (NLP),
possibly akin to code documentation.
Since natural language is, as the name suggests, ``natura'' to humans, NLP techniques can help make software analytics systems more explainable.
This suggests a \emph{dual system} that generates both prediction and NLP explanation \cite{hendricks2016generating}.
However, one should interpret the linguistic explanation with care because
we can derive explanation after-the-fact, and what is explained may not
necessarily reflect the true inner working of the model. A related technique is
to augment a deep network with interpretable \emph{structured knowledge}
such as rules, constraints or knowledge graphs \cite{hu2016deep}.

\subsection{Using external explainers}

This strategy keeps the original complex net intact but tries to
mimic or explain its behaviours using another interpretable model.
The ``mimic learning'' method, also known as \emph{knowledge distillation},
involves translating the predictive power of the complex net to
a simpler one (e.g., see \cite{hinton2015distilling}).
Distillation is highly useful in deep learning because neural networks are often redundant in neurons and connectivity.
For example, a popular training strategy known as dropout uses only 50\% of neurons at any
training step. A typical setup of knowledge distillation involves
first training a large ``teacher'' network, possibly an ensemble of networks, then using
the teacher network to guide the learning of a simpler student model.
The student model benefits from this setting due to knowledge of the domain already
distilled by the teacher network, which otherwise might be difficult
to learn directly from the raw data.

Another method aims at explaining the behaviours of the complex net
using an interpretable model. A successful case is LIME \cite{ribeiro2016should}, which generates
an explanation about a data instance locally. For example, given a sequence ``for(i=0; i < 10; i++)'', we can locally change the input to ``for(i=0; i < 10;)'' and observe the behaviour of the network. If
the behaviour changes, then we can expect that ``++'' is a relevant feature, even
if the network does not work directly on the space of tokens.

\section{Evaluation of explainability}

Prediction accuracy can be quantified into a number of widely-used measures such as precision, recall, F-measure, and Mean Absolute Errors. Explainability however refers to the qualitative aspect of a software analytics system, and thus it is difficult to measure explainability. We believe that for the explainable software analytics field to move forward, further research is needed to formalize the measures and methods for evaluating explainability of software analytics systems using evidence-based approaches. 
In this section, we sketch some potential directions for the evaluation of explainability.

A simple measure for explainability is the size of a model. The assumption here is that as the model grows in its size, it becomes harder for human engineers to understand it. The model size can be defined as the number of parameters (for linear regression models), the number of rules (for classification rule models), or the depth, the width, the number of edges, or the number of nodes (for decision trees or other graphical structure models) as used in some previous work (e.g. \cite{Wisam17}).

However, using the size of a model as an explainability indicator suffers from several limitations \cite{Freitas:2014:CCM}. First, the model's size captures only a syntactical representation of the model. It does not reflect the semantics of the model, which is an important factor in the model's explainability. In addition, previous experiments \cite{AllahyariL11} with 100 non-expert users found that larger decision trees and rule lists were in fact more understandable to users. Similar experiments can be done with software practitioners.

Conducting experiments with practitioners is the best way to evaluate if an software analytics model would work in practice. Those experiments would involve an analytics system working with software practitioners on a real task such as locating parts of a codebase that are relevant to a bug report. We would then evaluate the quality of explanations in the context of the task such as whether it leads to improved productivity or new ways of resolving a bug. Machine-produced explanations can be compared against explanations produced by human engineers when they assist their colleagues in trying to complete the same task. Other forms of experiments can involve: (i) users being asked to select a better explanation between two presented; (ii) users being provided with an explanation and an input, and then being asked to simulate the model's output; (iii) users being provided with an explanation, an input, and an output, and then being asked how to change the model in order to obtain to a desired output \cite{DoshiKim2017Interpretability}. Human study is however expensive and difficult to conduct properly.

\section{Discussion}

We have argued for explainable software analytics to facilitate human
understanding of machine prediction as a key to warrant trust from
software practitioners. There are other trust dimensions that also deserve attention.
Chief among them are \emph{reproducibility} (i.e., using the same code and data to
reproduce reported results), and \emph{repeatability} (i.e., applying the same method
to new data to achieve similar results). Further, should software analytics travels
the road of AI, then perhaps we want AI-based analytics agents that are not only
capable of explaining themselves but also aware of their own limitations.

\balance
\bibliographystyle{ACM-Reference-Format}


\end{document}